\documentclass[a4paper,fleqn]{cas-dc}
\usepackage[authoryear,longnamesfirst]{natbib}

\usepackage{float}
\usepackage{caption}
\usepackage{placeins}   
\usepackage{tcolorbox}
\usepackage{mdframed}
\renewcommand{\textfraction}{0.05}
\usepackage{seqsplit}
\usepackage{subcaption}

\begin{document}
\let\WriteBookmarks\relax

\renewcommand{\topfraction}{0.9}
\renewcommand{\bottomfraction}{0.8}
\renewcommand{\textfraction}{0.1}
\renewcommand{\floatpagefraction}{0.7}

\shorttitle{Dual-LLM Framework for Zero-Shot Cyber Threat Detection}

\author[1]{Abdullah Alghamdi}[orcid=0009-0008-7235-4488]\cormark[1]
\ead{as.alghamdi@uq.edu.au}

\author[1]{Siamak Layeghy}

\author[1]{Marius Portmann}

\affiliation[1]{organization={University of Queensland},
            city={Brisbane},
            country={Australia}}

\title[mode=title]{LLMs for Zero-Shot  Threat Detection via Structured Risk Indicators}

\cortext[1]{Corresponding author}


\begin{abstract}
We propose a two-stage large language model (LLM) framework for zero-shot detection of insider threats and advanced persistent threats (APTs) from heterogeneous security logs. The framework models user activity as chronological timelines and incorporates retrieval-augmented generation (RAG) to provide personalised behavioural context from each user's historical activity. Rather than performing end-to-end classification directly from raw logs, it first generates structured, interpretable sets of threat-specific risk indicators, which are then classified jointly across temporal sequences to capture attack patterns spanning multiple windows.
The framework is evaluated on two benchmark datasets, CERT r5.2 for insider threat detection and PicoDomain for APT detection, using four combinations of two open-weight LLMs under both retrieval and non-retrieval settings. All configurations outperform the previous state-of-the-art LLM-based framework (GABM), with the best configuration improving the F1-score by 11.40 percentage points on CERT r5.2 and 31.50 percentage points on PicoDomain. 
Results further show that retrieval mainly benefits weaker LLMs by generating more discriminative risk indicators, whereas stronger models achieve comparable performance without retrieved context. The most effective assignment of LLMs to the two stages depends on the dataset. These findings show that the quality of the generated risk indicators is the main driver of zero-shot cyber threat detection performance.

\end{abstract}




\maketitle

\section{Introduction}
Host-based intrusion detection systems (HIDS) play a critical role in modern cybersecurity by monitoring activity within an organisation's endpoints, including logon events, file access, device usage, and application-level activities such as email exchanges and host-recorded network interactions \citep{glassvanderlan2018survey}. Unlike network-based intrusion detection systems (NIDS), which analyse network traffic in transit, HIDS provide direct visibility into endpoint activity and are particularly effective for detecting insider threats and host-level stages of advanced persistent threats (APTs), including lateral movement, credential misuse, and data exfiltration \citep{prabhu2022primer, georgiadou2022detecting, laprade2020picodomain}. Because adversaries who obtain valid credentials, through phishing, credential theft, or insider collusion, can closely mimic legitimate users, endpoint-level detection often constitutes the last line of defence before significant damage occurs \citep{glasser2013bridging}. In practice, malicious behaviour is rarely confined to a single log source, and the same adversary may leave traces across endpoint and network logs. Effective detection therefore benefits from integrating these heterogeneous observations into a unified view of user behaviour.

Detecting such threats remains challenging because individual malicious actions frequently appear benign in isolation. Effective detection therefore requires contextual reasoning, including comparing current behaviour against historical baselines, correlating activity across heterogeneous log sources, and recognising attack stages that unfold over extended periods. Although both insider threats and APTs require contextual reasoning, the nature of that context differs substantially across threat classes. Insider threats typically emerge as gradual behavioural drift in user activity logs, such as unusual file access patterns, after-hours logons, or atypical communication behaviour \citep{glasser2013bridging}, whereas APT activity is reflected primarily through network-level patterns, including anomalous protocol sequences, command-and-control communication, and lateral movement \citep{mandiant_attack_lifecycle}.

Traditional machine learning approaches to intrusion detection, including Random Forest, Support Vector Machines, gradient-boosted trees \citep{alzaabi2024review}, and deep learning models such as recurrent and graph neural networks \citep{yuan2021deep}, have shown strong performance but depend on substantial labelled training data and often struggle to generalise to previously unseen attacks. Large language models (LLMs) have recently emerged as a promising alternative because they can analyse semi-structured security logs in a zero-shot setting without task-specific training \citep{kojima2022large}. However, current LLM-based approaches still perform end-to-end classification directly from raw logs, requiring the model to simultaneously interpret heterogeneous events, reason about user behaviour, and make a threat decision. The current state-of-the-art LLM framework, GABM \citep{ferraro2025generative}, illustrates this limitation by achieving perfect recall but low precision, resulting in a false-positive rate that limits practical deployment.

We hypothesise that this limitation arises not from the reasoning capability of LLMs themselves, but from the absence of an explicit behavioural abstraction between raw security logs and the final detection decision. Rather than asking an LLM to classify heterogeneous logs directly, we argue that it is more effective to first generate structured, interpretable sets of threat-specific risk indicators that summarise behavioural evidence, and then reason over their temporal evolution to detect multi-step attacks. We further hypothesise that grounding this process in each user's historical behaviour enables the generated risk indicators to better distinguish normal behavioural variation from genuine malicious activity while preserving the zero-shot setting.

To investigate these hypotheses, we propose a two-stage LLM framework for zero-shot cyber threat detection across heterogeneous security logs. The framework combines three complementary design principles. First, heterogeneous security logs are organised into chronological user timelines, providing a unified behavioural view across multiple log sources. Second, retrieval-augmented generation (RAG) provides personalised behavioural context by retrieving semantically similar historical activity from the same user. Third, LLMs generate structured, interpretable risk indicators that summarise each activity window before a second stage classifies their temporal evolution to identify attack patterns spanning multiple windows. To the best of our knowledge, combining personalised behavioural retrieval with LLM-generated structured risk indicators for zero-shot intrusion detection has not previously been investigated.

We evaluate the proposed framework on two complementary benchmark datasets: CERT r5.2 \citep{glasser2013bridging}, representing insider threat detection from host activity logs, and PicoDomain \citep{laprade2020picodomain}, representing APT detection from Zeek network logs. Across four combinations of two open-weight LLMs, evaluated with and without retrieval augmentation, every configuration outperforms the previous state-of-the-art LLM-based framework (GABM). The best configuration improves the F1-score by 11.40 percentage points on CERT r5.2 and 31.50 percentage points on PicoDomain, while substantially improving precision without sacrificing high recall. Beyond these performance gains, the evaluation provides new insights into how retrieval augmentation and model capability interact during zero-shot threat detection.

The main contributions of this work are as follows.

\begin{itemize}
    \item We propose a zero-shot LLM framework that replaces direct classification of heterogeneous security logs with structured risk-indicator generation followed by temporal classification, providing a unified approach to insider threat and APT detection.

    \item We present a comprehensive empirical analysis of retrieval augmentation and model capability, showing that retrieval primarily benefits weaker LLMs by improving the quality of generated risk indicators, while the most effective assignment of LLMs to the two stages depends on the characteristics of the dataset.
\end{itemize}

The remainder of this paper is organised as follows. Section~\ref{sec:related} reviews related work on traditional, deep learning, and LLM-based approaches to intrusion detection. Section~\ref{sec:methodology} presents the proposed methodology. Section~\ref{sec:experimental_setup} describes the experimental setup. Section~\ref{sec:results} presents the experimental results and ablation studies. Finally, Section~\ref{sec:conclusion} concludes the paper and outlines future research directions.

\section{Related Work} \label{sec:related}

Recent advances in LLMs have shifted cybersecurity log analysis from supervised learning towards zero-shot reasoning over semi-structured security data. This section reviews existing LLM-based approaches to cyber threat detection and RAG, and positions the proposed work within this literature.

\subsection{Large Language Models for Cyber Threat Detection}

Large language models have recently emerged as a promising alternative to traditional supervised intrusion detection by enabling zero-shot reasoning over heterogeneous security logs without task-specific training \citep{kojima2022large}. Existing approaches explore this direction through prompt engineering \citep{li2025redchronos}, parameter-efficient fine-tuning \citep{kong2025dmfi,song2025confront}, and multi-view behavioural modelling \citep{song2025insightllm}. A broader survey by \citet{xu2024llmsurvey} identifies low precision under high-volume telemetry as one of the principal challenges facing LLM-based cybersecurity systems.

Despite their methodological differences, most existing approaches rely on end-to-end classification directly from raw or lightly processed log data. This requires the LLM to simultaneously interpret heterogeneous security events, reason about user behaviour, and make a threat decision within a single inference step. Our preliminary experiments (Section~\ref{sec:direct_llm}) show that this design leads to unstable predictions and low precision, consistent with the observations reported by \citet{xu2024llmsurvey}. A separate line of research therefore limits the role of LLMs to post-hoc investigation and explanation rather than detection itself. For example, eX-NIDS \citep{houssel2026exnids} employs LLMs to explain alerts generated by an external intrusion detection system, while \citet{pletzer2026anomaly} use LLMs to analyse candidate attacks identified by a statistical anomaly detector. Although these approaches improve interpretability, the LLM no longer contributes directly to the detection process.

The closest LLM-based detection frameworks to this work are GABM \citep{ferraro2025generative} and Audit-LLM \citep{song2024audit}, both of which decompose detection into multiple cooperating LLM agents. GABM represents the current state of the art on the benchmark datasets considered in this work, but achieves high recall at the expense of low precision. Audit-LLM similarly employs multiple agents to coordinate threat analysis but reports user-level evaluation metrics, preventing direct comparison with window-level detection. Although these frameworks demonstrate the potential of collaborative LLM reasoning, they continue to exchange free-form natural-language reasoning between stages rather than structured behavioural evidence, and neither explicitly models user behaviour relative to a personalised historical baseline.

\subsection{Retrieval-Augmented Generation for Cybersecurity}

Retrieval-Augmented Generation  extends LLM reasoning by incorporating externally retrieved information \citep{lewis2020rag}. Existing cybersecurity applications primarily retrieve threat intelligence from external knowledge sources, such as MITRE ATT\&CK, CVE repositories, and vendor advisories, to improve reasoning about emerging attacks \citep{paul2025llmrag,borah2025ragsec}. In these systems, retrieval enriches the model with domain knowledge that is not contained within its parameters.

Our use of retrieval addresses a different problem. Instead of retrieving external threat intelligence, retrieval provides personalised behavioural context by comparing each activity window against the same user's historical behaviour. This shifts the role of retrieval from knowledge augmentation to behavioural grounding, enabling risk indicators to be generated relative to an individual's established activity patterns rather than generic notions of suspicious behaviour. To the best of our knowledge, combining personalised behavioural retrieval with structured LLM-generated risk indicators for zero-shot cyber threat detection has not previously been investigated.


\section{Methodology}\label{sec:methodology}
 
\begin{figure*}[!t]
    \centering
     \vspace{-1mm}
\includegraphics[width=\textwidth, trim=0 490 0 140, clip]{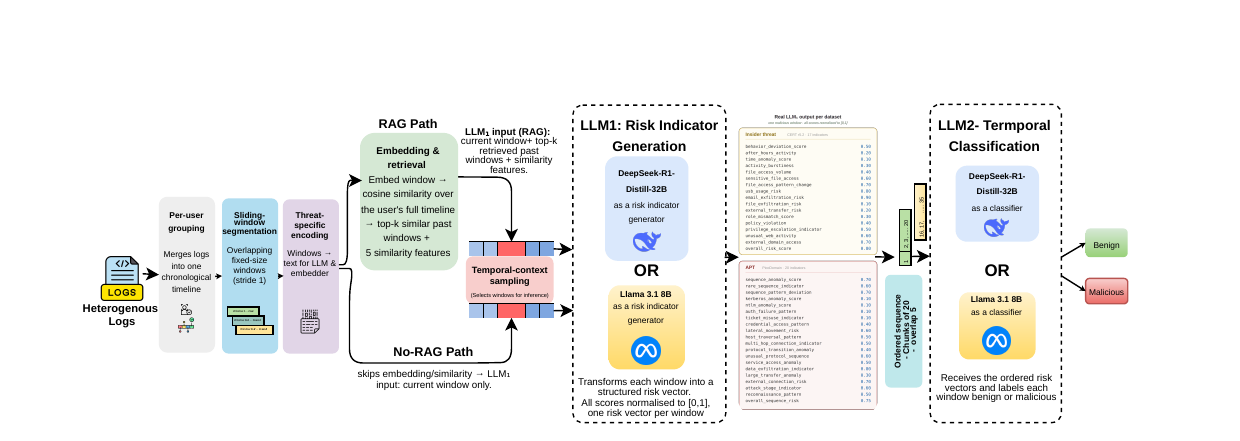}  

\caption{Overview of the proposed retrieval-augmented dual-LLM framework. Heterogeneous logs are grouped into per-user
chronological timelines, segmented into overlapping windows, and
encoded into a threat-specific textual representation. Under the
retrieval-augmented (RAG) configuration, each window is embedded
and compared against the user's full history to retrieve the
top-$k$ most similar past windows and five similarity features;
the No-RAG configuration omits this step. LLM\textsubscript{1}
generates a threat-specific structured feature vector per window
($17$ insider-threat or $20$ APT indicators), and
LLM\textsubscript{2} classifies the resulting ordered feature
sequences as benign or malicious. Both stages are instantiated
with DeepSeek-R1-Distill-Qwen 32B or Llama~3.1 8B, yielding four
model combinations under each retrieval setting.}
    \label{fig:pipeline}
\end{figure*}

The proposed framework consists of three main components:
(i)~data preparation and temporal windowing,
(ii)~behaviour-aware retrieval for contextual augmentation, and
(iii)~dual-LLM inference for feature generation and
classification over ordered windows.
Given the scale of the windowed datasets, a deterministic
temporal-context sampling step (Section~\ref{sec:sampling})
selects the windows that undergo LLM inference.
We evaluate the framework under two retrieval configurations, RAG and No-RAG, to isolate the contribution of retrieval across both models. 
Figure~\ref{fig:pipeline} provides an overview of the pipeline.
The same architectural backbone is applied to both datasets
considered in this work, with the windowing, window
representation, and LLM\textsubscript{1} feature set
instantiated according to the threat type.

\subsection{Data Preparation and Temporal Windowing}
\label{sec:data_preparation}

\subsubsection{Per-user log grouping.}
We group by user because malicious behaviour typically spans several
log sources rather than appearing within any single one: an insider's
file access, logon timing, and email activity (or an APT actor's
beaconing and lateral movement) form a recognisable pattern only
when a user's activity is viewed as a whole. Grouping all of a user's
records into one chronological timeline therefore gives each
inference window the full context of that user's behaviour, rather
than a fragment confined to a single log type or connection.

Concretely, logs are grouped by user and sorted chronologically. For
PicoDomain, each log entry is associated with a user through the
host-to-user mapping provided in the dataset ground truth
(e.g.\ \texttt{BDUCK}, \texttt{JDOE}). For CERT~r5.2, the five log
sources used in this work (logon, device, file, email, HTTP) are
merged into a single chronological activity sequence per user.

\subsubsection{Sliding-window construction.}
Each user's chronologically sorted log sequence is segmented into
sliding windows of fixed size $w$ with stride~1, where $w{=}10$
for CERT~r5.2 and $w{=}5$ for PicoDomain.
A window is labelled malicious if any of its constituent log
entries carries a malicious label; otherwise it is labelled benign.
Under stride-1 windowing, a single malicious log entry propagates
across up to $w$ consecutive windows, all receiving a malicious
label; this any-positive labelling rule is applied consistently
across all experimental conditions, ensuring that observed
performance differences are attributable to pipeline configuration
rather than labelling variation.
The larger window for CERT~r5.2 reflects the gradual, multi-step
nature of insider threat behaviour, where a meaningful activity
sequence spans multiple log types; the smaller window for
PicoDomain is sufficient given that APT activity manifests in
short, concentrated bursts.

Each window is encoded for two purposes. For the embedding model
(Section~\ref{sec:rag}), the window is represented as a compact
protocol-and-destination token sequence such as
\texttt{http:internal $\rightarrow$ ssl:c2 $\rightarrow$ dns:external
$\rightarrow$ kerberos: internal $\rightarrow$ conn:external}, where
destinations are categorised as \texttt{internal} (10.99.99.0/24),
\texttt{c2} (known C2 addresses 3.3.3.5 and 1.1.1.11), or
\texttt{external}.

LLM\textsubscript{1} (Section~\ref{sec:llm1}) receives a richer
representation: the full Zeek log records for the window across all
log types, accompanied by this protocol-destination sequence as a
compact structural summary.
For CERT~r5.2, each event in the window is rendered as a structured
event object: a JSON record retaining the threat-relevant fields for
its activity type (action and PC for logon, recipients and attachment
metadata for email, filename and removable-media flag for file, URL
for HTTP, and file-tree metadata for device), with events preserved
in temporal order. This representation serves both the embedding model
and LLM\textsubscript{1}. After windowing, PicoDomain yields
460{,}033 windows from 537{,}840 raw log entries, while CERT~r5.2
yields 3{,}400{,}220 windows from the logs of its 99 malicious users,
drawn from a full corpus of approximately 79.9~million entries
(Table~\ref{tab:dataset}).

\begin{table}[!t]
\centering
\caption{Dataset overview and sampling summary. For CERT~r5.2,
windowing and all subsequent stages are applied to the 99 malicious
users' logs; \emph{Raw entries} reports the full corpus. Benign
windows are therefore drawn from the benign activity within these
users' own timelines.}
\label{tab:dataset}
\resizebox{\columnwidth}{!}{%
\begin{tabular}{lrrrrrr}
\toprule
Dataset & Raw entries & Users (Total) & Mal. users & Windows & Sampled & Mal. / Ben. \\
\midrule
PicoDomain & 537{,}840 & 7 & 5 & 460{,}033 & 2{,}741 & 1{,}000 / 1{,}741 \\
CERT r5.2 & $\sim$79.9\,M & 2{,}000 & 99 & 3{,}400{,}220 
& 9{,}571 & 4{,}410 / 5{,}161 \\

\bottomrule
\end{tabular}%
}
\end{table}

\subsection{Behaviour-Aware Retrieval for Contextual
            Augmentation}
\label{sec:rag}

We incorporate a retrieval-augmented generation (RAG)
mechanism to provide each window with behavioural context
drawn from the same user's history.
For each window, the most similar past windows are retrieved
and attached as context for LLM\textsubscript{1}, enabling
comparison of current behaviour against a personalised
historical baseline.

\subsubsection{Window embedding.}
Each window's textual representation is encoded into a 384-dimensional dense vector using
\texttt{all-MiniLM-L6-v2} ~\citep{reimers2019sentencebert},
a lightweight sentence transformer model.
Embeddings are computed once per window and stored for
reuse across all downstream steps.

\subsubsection{Similarity computation and features.}
For each window~$i$, cosine similarity is computed against all prior
windows within a lookback horizon of 500~windows from the same user.
Five deterministic similarity features are derived from the resulting
similarity distribution: \textit{sim\_max} (maximum similarity to any
past window), \textit{sim\_topk\_avg} (mean similarity over the top
five matches), \textit{sim\_topk\_std} (standard deviation over the
top five matches), \textit{sim\_global\_mean} (mean similarity over
all windows in the lookback horizon), and \textit{sim\_gap}
(difference between \textit{sim\_max} and \textit{sim\_topk\_avg}).
These features provide a quantitative signal of how anomalous the
current window is relative to the user's recent history, and are
passed to LLM\textsubscript{1} alongside the log content.

\subsubsection{Retrieval.}
The top $k{=}3$ most similar past windows, by cosine
similarity over the embeddings, are retrieved from the
preceding 500-window lookback horizon within the same
user's history and attached as context for
LLM\textsubscript{1}.
For CERT~r5.2, each retrieved window contains the structured
event fields of its constituent log entries.
For PicoDomain, each retrieved window contains the full Zeek
log entries across all log types, with unique connection
identifiers and sensor names removed.
The first window of each user's sequence has no retrievable
history and is processed with an empty context.
Labels are removed from all retrieved windows before they are
presented to LLM\textsubscript{1}, preventing label leakage.

\subsubsection{No-RAG configuration.}
\label{sec:norag}
To isolate the contribution of retrieval, we define a No-RAG
configuration in which LLM\textsubscript{1} receives only the current
window's log content, with no retrieved past windows and no
deterministic similarity features. The LLM\textsubscript{1} prompt is
trimmed accordingly: the instruction to compare the current window
against past windows is removed, and the model must generate risk
indicators from the current window alone. All other pipeline
components remain identical; the No-RAG configuration differs from the
RAG configuration solely in the removal of the retrieval mechanism: the retrieved past windows and the similarity features derived from
them.


\subsubsection{Temporal-context sampling.}
\label{sec:sampling}
Given the large scale of the windowed datasets, applying LLM
inference to all windows is computationally impractical.
We therefore adopt a sampling strategy that preserves malicious activity while retaining sufficient benign context
for temporal analysis.

Because embedding, similarity computation, and retrieval
(Section~\ref{sec:rag}) are performed over each user's
complete windowed timeline, sampling determines only which
windows undergo LLM inference; retrieved context is therefore
drawn from the user's full history rather than from the
sampled subset.

For each user, contiguous sequences of malicious windows
(i.e.\ \emph{malicious bursts}) are identified.
A symmetric context of benign windows is retained before and
after each burst, preserving the temporal transition between
normal and anomalous behaviour.
For PicoDomain, all malicious bursts are retained with a
context of $\pm 10$ benign windows, yielding 2{,}741~sampled
windows.
For CERT~r5.2, the largest malicious burst per user is selected with
a context of $\pm 30$ benign windows, yielding 9{,}571~sampled
windows across 99~users; selecting the largest burst bounds the
sampled set while capturing each user's principal malicious episode
together with its surrounding normal behaviour. The larger context
for CERT~r5.2 reflects the gradual nature of insider threat
behaviour, where the transition between normal and anomalous activity
spans many windows; the smaller context for PicoDomain is sufficient
given the concentrated, abrupt nature of APT beaconing bursts.
 
The sampling procedure is deterministic and depends only on
window labels and sequential positions.
This ensures that the exact same subset of windows is evaluated
across all experimental conditions (including all
second-stage classifier variants) so that observed
performance differences are attributable solely to the pipeline
configuration rather than to variation in the evaluation data.
The resulting class distributions are summarised in
Table~\ref{tab:dataset}.

Because sampling concentrates on malicious bursts and their immediate
context, the reported precision and F1 reflect this evaluation
distribution rather than the full-corpus base rate. As the sampled
subset is identical across all conditions, this does not affect the
relative comparisons that are the focus of this work.

\subsection{Feature Generation (LLM\textsubscript{1})}
\label{sec:llm1}

The proposed framework employs a two-stage inference process
in a zero-shot setting. Its central component is
LLM\textsubscript{1}, a threat-aware feature generator that
transforms each raw log window into a structured vector of
normalised risk indicators, replacing the hand-engineered
feature extractors used in conventional pipelines. The second
stage, LLM\textsubscript{2}, is a downstream sequence
classifier that reasons over the ordered feature vectors
LLM\textsubscript{1} produces. The framework instantiates
both stages using two open-weight LLMs of differing
capability (DeepSeek-R1-Distill-Qwen 32B and Llama~3.1
8B), both quantised to 4-bit precision (Q4\_K\_M) and
deployed via \texttt{llama.cpp}. 

\begin{figure}
    \centering
    \includegraphics[width=\columnwidth]{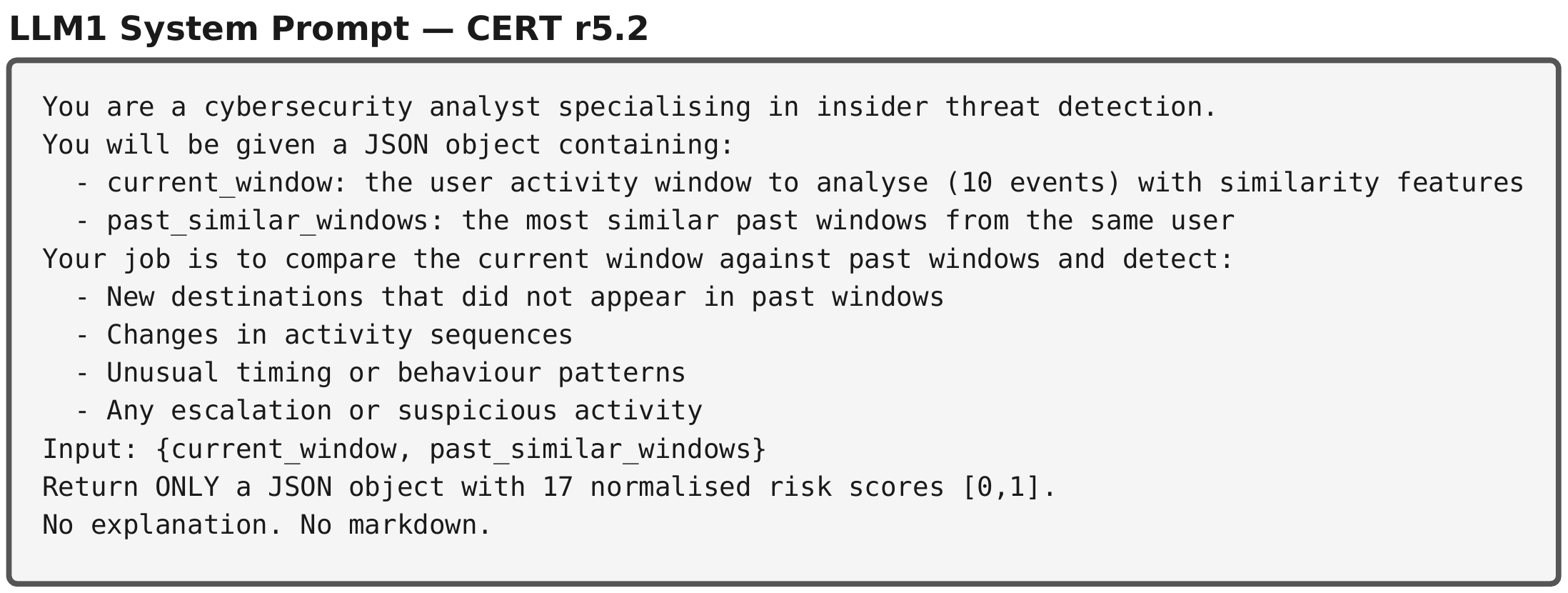}
    \includegraphics[width=\columnwidth]{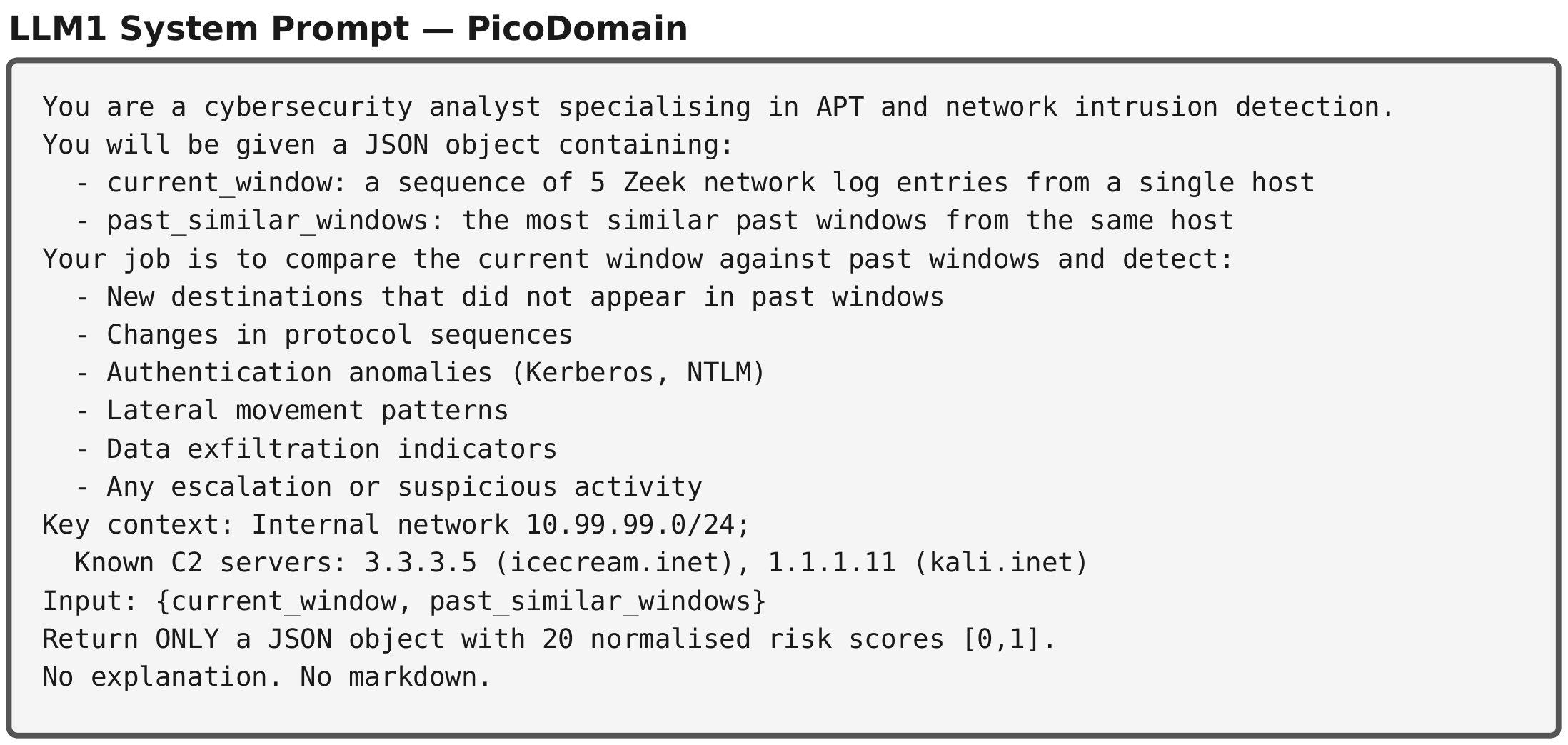}
    \caption{LLM\textsubscript{1} system prompts:
             CERT~r5.2 (top) and PicoDomain (bottom).}
    \label{fig:prompt-llm1}
\end{figure}

LLM\textsubscript{1} transforms each sampled window into a
structured feature representation: a set of normalised risk
indicators tailored to the threat type of each dataset. Its
core input is the current window's log content (with labels
removed). In the retrieval-augmented configuration,
LLM\textsubscript{1} additionally receives up to three retrieved
past windows from the same user together with the deterministic
similarity features defined in Section~\ref{sec:rag}, which
quantify how anomalous the current window is relative to the
user's recent history; the contribution of this retrieved
context is analysed in Section~\ref{sec:overall}
(Table~\ref{tab:cross_model}).
In this configuration, the prompt instructs the model to act
as a cybersecurity analyst and to compare the current window
against the provided past windows, identifying deviations in
destinations, protocol or activity sequences, timing patterns, and any escalation of suspicious behaviour.
In the No-RAG configuration (Section~\ref{sec:norag}), the
current window's log content is the model's sole input, and
the prompt omits the comparison instruction, requiring the
risk indicators to be generated from the window alone.
The model returns a JSON object containing normalised risk scores, each bounded to [0,1]. The discriminative quality of the generated features is assessed using SHAP analysis, reported in Section~\ref{sec:feature_quality}. The full set of risk indicators is listed in Tables~\ref{tab:features_cert} and~\ref{tab:features_pico}, and the LLM\textsubscript{1} and LLM\textsubscript{2} system prompts are shown in Figures~\ref{fig:prompt-llm1} and~\ref{fig:prompt-llm2}.

\paragraph{Dataset-specific feature sets.}
The feature sets are tailored to the threat type of each
dataset. For each dataset, the feature set is constructed to
satisfy two constraints: (i)~\emph{coverage}: every
behavioural dimension represented in the dataset's documented
threat scenarios must be observable through at least one
feature; and (ii)~\emph{extractability}: each feature must
be derivable from the log types the dataset actually provides. Scenario documentation is used only to define the feature schema (analogous to the threat-model knowledge that informs detection rules in operational deployments); no ground-truth labels inform feature design, and the same fixed feature set is applied uniformly across all users and experimental configurations.
For CERT~r5.2, the 17~features capture behavioural indicators
of insider threat across the scenarios documented in the
dataset (data leakage, intellectual property theft, and IT
sabotage), organised into seven behavioural categories:
deviation from the user's behavioural baseline, temporal
anomalies, data access, data exfiltration across removable
media, email, and file channels, privilege misuse, network
activity, and a single aggregate risk summary. Each feature is
extractable from the five CERT log types used in this work
(logon, device, file, email, and HTTP), aligning the feature
set with both MITRE ATT\&CK tactics~\citep{mitre2026attack} and
User and Entity Behaviour Analytics (UEBA) indicators. SHAP
analysis (Section~\ref{sec:feature_quality}) shows that both
models rank file-access and exfiltration indicators most
highly, though they differ in the single dominant feature; the
relative importance reflects the attack characteristics of the
dataset, with USB-based exfiltration scenarios elevating
\textit{usb\_usage\_risk} for the stronger model.

For PicoDomain, the 20~features target network-level indicators
of APT activity aligned with the stages of the APT kill chain,
organised into seven categories: sequence-level
anomalies in the host's connection patterns, authentication
abuse (Kerberos and NTLM), lateral movement between internal
hosts, protocol-level irregularities, data exfiltration,
attack-stage progression, and a single aggregate risk summary.
Each feature is extractable from the Zeek log types in the
dataset (CONN, DNS, DCE\_RPC, SSL, and Kerberos). Unlike
CERT~r5.2, individual-window feature signal on PicoDomain is
inherently weak (Section~\ref{sec:feature_quality}), a
consequence of the stealthy nature of APT beaconing rather than
feature design; the feature set nonetheless provides a consistent structured representation for LLM\textsubscript{2} to classify.
This dataset-specific design allows LLM\textsubscript{1} to
focus on the most discriminative signals for each threat type
rather than relying on a generic feature set.

LLM\textsubscript{1} receives a JSON object containing the current
activity window and its retrieved similar past windows, and returns a
fixed-length vector of risk scores corresponding to the features in
Tables~\ref{tab:features_cert} and~\ref{tab:features_pico}.

\begin{table*}[!t]
\centering
\caption{LLM\textsubscript{1} risk indicators for PicoDomain,
grouped by the category of behaviour they characterise. All
scores are normalised to $[0,1]$.}
\label{tab:features_pico}
\scriptsize
\setlength{\tabcolsep}{5pt}
\renewcommand{\arraystretch}{1.1}
\begin{tabular}{@{}p{4.2cm}p{\dimexpr\linewidth-4.2cm-2\tabcolsep\relax}@{}}
\toprule
Feature & Description \\
\midrule
\multicolumn{2}{@{}l}{\textit{Sequence}}\\
\seqsplit{sequence\_anomaly\_score} &
Overall degree to which the window's protocol and connection
sequence departs from the user's established patterns. \\
\seqsplit{rare\_sequence\_indicator} &
Presence of protocol or connection sequences rarely or never
seen in the host's prior history, flagging novel network
behaviour. \\
\seqsplit{sequence\_pattern\_deviation} &
Extent of deviation in the ordering of events relative to the
host's normal sequential behaviour. \\
\midrule
\multicolumn{2}{@{}l}{\textit{Authentication}}\\
\seqsplit{kerberos\_anomaly\_score} &
Anomalous Kerberos authentication activity, such as unusual
ticket-granting requests. Kerberos abuse is a common signature
of credential-based attacks in Windows domains. \\
\seqsplit{ntlm\_anomaly\_score} &
Suspicious NTLM authentication activity inconsistent with the
host's normal authentication profile. \\
\seqsplit{auth\_failure\_pattern} &
Patterns of repeated or distributed authentication failures,
indicative of credential probing or brute-force attempts. \\
\seqsplit{ticket\_misuse\_indicator} &
Evidence of Kerberos ticket abuse or reuse, a signature of
attacks such as pass-the-ticket. \\
\seqsplit{credential\_access\_pattern} &
Suspicious patterns of credential access or harvesting,
capturing attempts to obtain account secrets. \\
\midrule
\multicolumn{2}{@{}l}{\textit{Lateral movement}}\\
\seqsplit{lateral\_movement\_risk} &
Risk of movement between internal hosts, a core stage of APT
campaigns as adversaries pivot toward their objective. \\
\seqsplit{host\_traversal\_pattern} &
Patterns of connections traversing multiple internal hosts
beyond the host's normal communication footprint. \\
\seqsplit{multi\_hop\_connection\_indicator} &
Presence of multi-hop connection chains characteristic of
pivoting through a network. \\
\midrule
\multicolumn{2}{@{}l}{\textit{Protocol}}\\
\seqsplit{protocol\_transition\_anomaly} &
Unusual transitions between network protocols within a
connection sequence, which can indicate covert channels. \\
\seqsplit{unusual\_protocol\_sequence} &
Protocol orderings within the window that diverge from
expected service behaviour. \\
\seqsplit{service\_access\_anomaly} &
Anomalous access to network services relative to the host's
normal service-usage profile, capturing service enumeration. \\
\midrule
\multicolumn{2}{@{}l}{\textit{Exfiltration}}\\
\seqsplit{data\_exfiltration\_indicator} &
Signs that data is being transferred outside the network
boundary, the defining objective of an APT campaign. \\
\seqsplit{large\_transfer\_anomaly} &
Unusually large data transfers inconsistent with the host's
normal traffic volume. \\
\seqsplit{external\_connection\_risk} &
Risk associated with connections to external or
command-and-control destinations. \\
\midrule
\multicolumn{2}{@{}l}{\textit{Attack progression}}\\
\seqsplit{attack\_stage\_indicator} &
Evidence positioning the window within a recognised stage of
an APT campaign, from reconnaissance through to exfiltration. \\
\seqsplit{reconnaissance\_pattern} &
Patterns of port scanning or service discovery, characteristic
of the reconnaissance stage of an attack. \\
\midrule
\multicolumn{2}{@{}l}{\textit{Aggregate}}\\
\seqsplit{overall\_sequence\_risk} &
Holistic risk assessment for the window, integrating all
preceding indicators into a single summary signal. \\
\bottomrule
\end{tabular}
\end{table*}


\begin{table*}[!t]
\centering
\caption{LLM\textsubscript{1} risk indicators for CERT~r5.2,
grouped by the category of behaviour they characterise. All
scores are normalised to $[0,1]$.}
\label{tab:features_cert}
\scriptsize
\setlength{\tabcolsep}{5pt}
\renewcommand{\arraystretch}{1.1}
\begin{tabular}{@{}p{4.2cm}p{\dimexpr\linewidth-4.2cm-2\tabcolsep\relax}@{}}
\toprule
Feature & Description \\
\midrule
\multicolumn{2}{@{}l}{\textit{Behavioural baseline}}\\
\seqsplit{behavior\_deviation\_score} &
Overall degree to which the window departs from the user's
established activity profile. Insider threats are fundamentally
deviations from a user's own behavioural baseline. \\
\midrule
\multicolumn{2}{@{}l}{\textit{Temporal}}\\
\seqsplit{after\_hours\_activity} &
Degree to which actions fall outside the user's normal working
hours. Insider data theft is frequently timed to off-hours to
reduce the chance of observation. \\
\seqsplit{time\_anomaly\_score} &
Irregularity of event timing relative to the user's historical
rhythm, capturing atypical intervals and weekend activity not
explained by working hours alone. \\
\seqsplit{activity\_burstiness} &
Concentration of many actions into a short interval. Bulk data
collection or staging produces dense activity bursts atypical
of routine work. \\
\midrule
\multicolumn{2}{@{}l}{\textit{Data access}}\\
\seqsplit{file\_access\_volume} &
Volume of file-access operations relative to the user's norm.
An elevated access count is a common precursor to large-scale
exfiltration. \\
\seqsplit{sensitive\_file\_access} &
Degree of interaction with sensitive, restricted, or
high-value files, flagging direct contact with the assets most
likely to be targeted. \\
\seqsplit{file\_access\_pattern\_change} &
Shift in the type, location, or breadth of files accessed
relative to prior behaviour, indicating a move from
role-consistent to anomalous access. \\
\midrule
\multicolumn{2}{@{}l}{\textit{Data exfiltration}}\\
\seqsplit{usb\_usage\_risk} &
Risk of removable-media operations, scaled by the frequency of
device use and the sensitivity of files transferred. Removable
storage is the principal physical exfiltration channel in
CERT~r5.2. \\
\seqsplit{email\_exfiltration\_risk} &
Likelihood that email activity constitutes data leakage, such
as large or unusual attachments sent to external addresses. \\
\seqsplit{file\_exfiltration\_risk} &
Likelihood that file operations represent exfiltration, such
as copying sensitive files toward staging or external
destinations. \\
\seqsplit{external\_transfer\_risk} &
Risk of data crossing the organisational boundary to external
recipients or systems, aggregating signals across email, web,
and file channels. \\
\midrule
\multicolumn{2}{@{}l}{\textit{Privilege misuse}}\\
\seqsplit{role\_mismatch\_score} &
Extent to which observed actions diverge from those typical of
the user's job role. Insider misuse often surfaces as access
unrelated to legitimate responsibilities. \\
\seqsplit{policy\_violation} &
Presence of actions that contravene organisational security
policy, a direct indicator of intentional misconduct. \\
\seqsplit{privilege\_escalation\_indicator} &
Evidence of attempts to obtain or exercise access beyond the
user's authorised permissions. \\
\midrule
\multicolumn{2}{@{}l}{\textit{Network}}\\
\seqsplit{unusual\_web\_activity} &
Anomaly in web-browsing behaviour relative to the user's norm,
capturing reconnaissance or unsanctioned upload activity. \\
\seqsplit{external\_domain\_access} &
Degree of connection to external or previously unseen domains,
flagging communication with untrusted destinations. \\
\midrule
\multicolumn{2}{@{}l}{\textit{Aggregate}}\\
\seqsplit{overall\_risk\_score} &
Holistic risk assessment for the window, integrating all
preceding indicators into a single summary signal. \\
\bottomrule
\end{tabular}
\end{table*}


\subsection{Sequence Classification (LLM\textsubscript{2})}
\label{sec:llm2}

\begin{figure}
    \centering
    \includegraphics[width=\columnwidth]{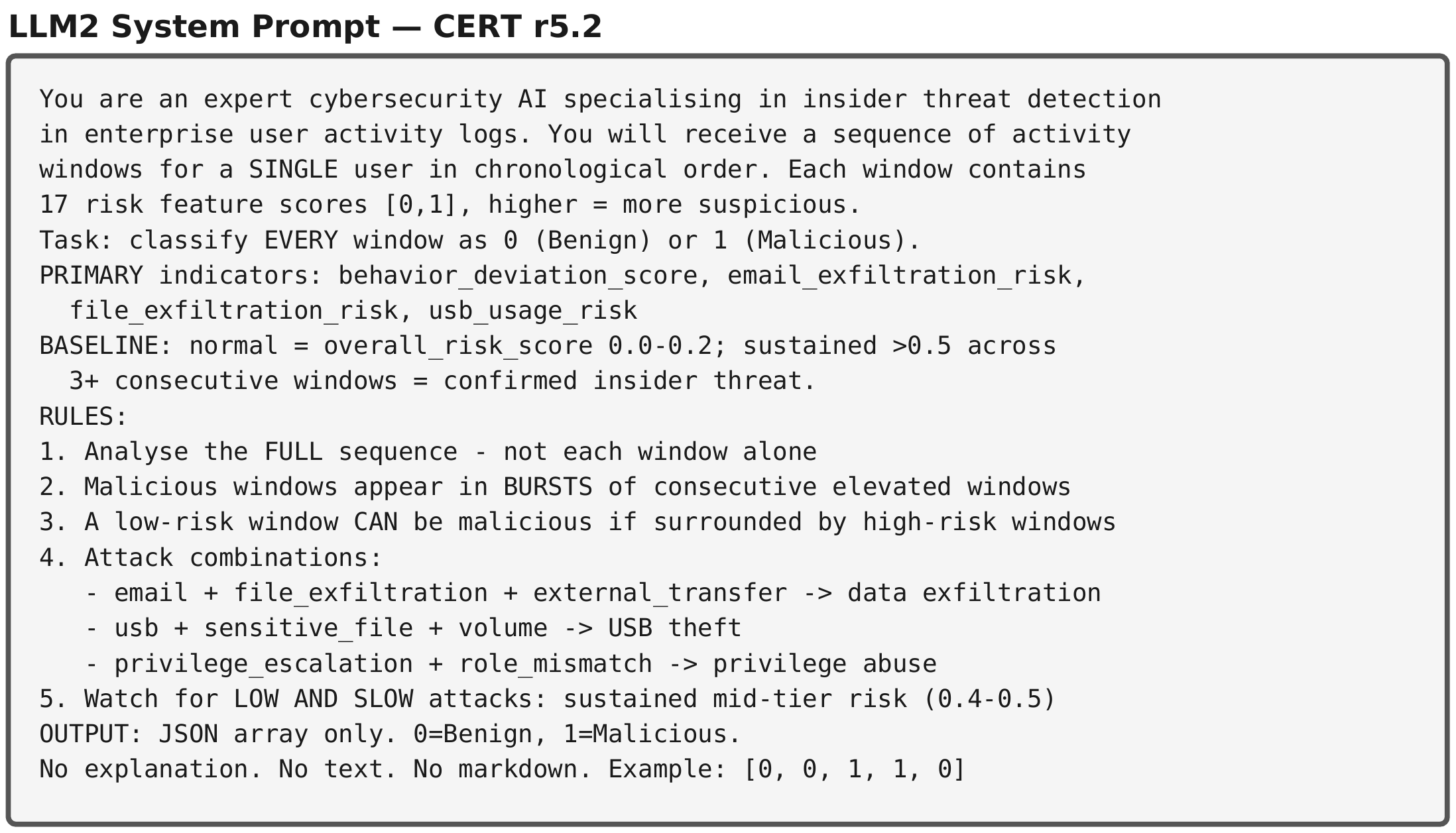}
    \includegraphics[width=\columnwidth]{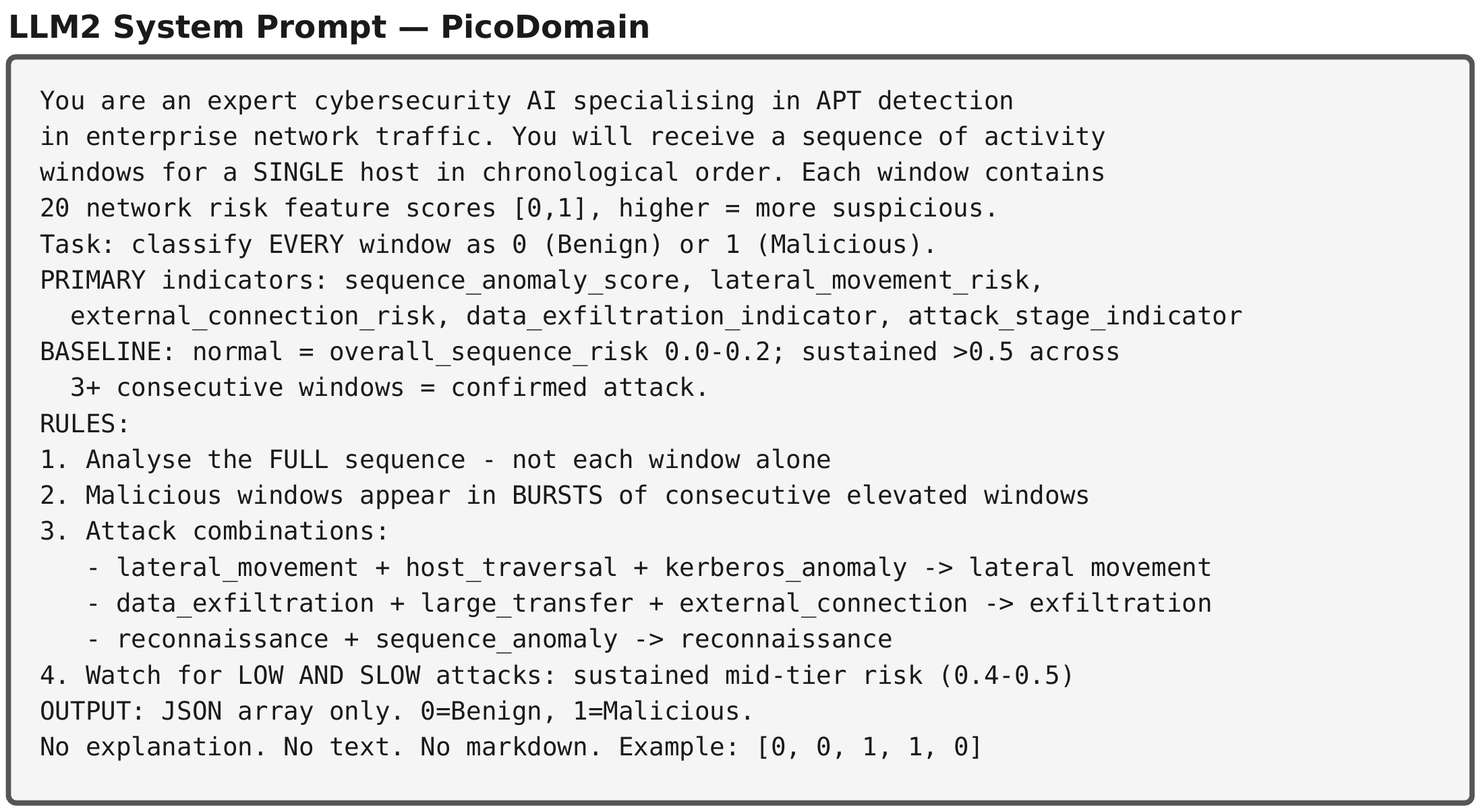}
    \caption{LLM\textsubscript{2} system prompts:
             CERT~r5.2 (top) and PicoDomain (bottom).}
    \label{fig:prompt-llm2}
\end{figure}
 
LLM\textsubscript{2} receives the unlabelled feature representations
produced by LLM\textsubscript{1} for a given user, arranged in
chronological order, and produces a binary classification
(benign or malicious) for each window.
In both configurations, LLM\textsubscript{2} receives only the
risk-indicator vectors produced by LLM\textsubscript{1}, never raw log content; the deterministic similarity features of
Section~\ref{sec:rag} serve solely as input context for feature
generation and are not propagated downstream. LLM\textsubscript{2}
is identical across the RAG and No-RAG configurations, so
performance differences between the two settings are
attributable solely to the features LLM\textsubscript{1}
produces.
 
\paragraph{Chunked processing.}
Due to context-window constraints, the ordered feature
sequence for each user is divided into chunks of 20~windows
with an overlap of 5~windows between consecutive chunks.
The overlap ensures that burst boundaries are not missed at
chunk edges.
When a window appears in multiple chunks, the later chunk's
prediction is retained, since there the window has its subsequent
context in view.

\paragraph{Burst-aware reasoning.}
The prompt instructs LLM\textsubscript{2} to reason over the
full temporal sequence rather than evaluating each window in
isolation. The model identifies regions of consecutive windows
with elevated risk scores, recognises coordinated attack
patterns such as data exfiltration, privilege abuse, and
suspicious access based on related feature combinations
(pattern descriptions instantiated per dataset, Figure~\ref{fig:prompt-llm2}), and uses surrounding context to classify ambiguous windows that may appear benign in isolation but fall within a broader malicious burst. This prompt design is motivated by the tendency of malicious activity in both APT and insider threat scenarios to manifest as bursts of sustained elevated risk rather than isolated anomalous windows.

LLM\textsubscript{2} receives a chronological sequence of risk-score
vectors produced by LLM\textsubscript{1} for a single user and assigns
a binary label to every window ($0$=benign, $1$=malicious).

 
\section{Experimental Setup}\label{sec:experimental_setup}
 
\subsection{Datasets}
\label{sec:datasets}
 
We evaluate the proposed framework on two publicly available
cybersecurity datasets: PicoDomain and CERT~r5.2.
 
\paragraph{PicoDomain.}
The PicoDomain~\citep{laprade2020picodomain} is a high-fidelity
Zeek network log dataset simulating an enterprise environment
under an advanced persistent threat (APT) campaign.
The dataset covers a three-day period and contains 537{,}840
log rows across multiple Zeek log types, including CONN, DNS,
DCE\_RPC, SSL, and Kerberos.
Logs are associated with users based on the host-to-user
mapping provided in the dataset ground truth.
Following preprocessing, the dataset contains 7 users, of which 2 exhibit exclusively benign activity and are excluded, leaving 5 users (all associated with malicious activity) for evaluation.
The ground truth contains 80 red team events, of which 79
correspond to malicious actions; the remaining event marks
the end of the attack campaign and is not considered a
malicious activity in this work.
Labels are derived by matching log entries against the 79
malicious events using a temporal window of $\pm 5$~seconds
and IP address matching.

\paragraph{CERT~r5.2.}
The CERT Insider Threat Dataset
r5.2~\citep{glasser2013bridging} is a synthetic dataset
capturing user activity logs for a simulated organisation
over an 18-month period.
It contains 2{,}000 employees, of whom 99 are associated with
labelled malicious activity across four insider threat
scenarios (data leakage, two variants of intellectual property
theft, and IT sabotage).
The dataset provides seven log types, of which five are used
in this work: logon, device, file, email, and HTTP.
The remaining two (decoy file access and psychometric data)
are excluded as they do not represent genuine user activity.
Labels are derived by matching event identifiers against the
ground truth from scenario folders.

\subsection{Implementation Details}
\label{sec:implementation}

DeepSeek-R1-Distill-Qwen-32B is run locally via
\texttt{llama.\allowbreak cpp} on NVIDIA A100 and L40S GPU nodes of
the Bunya HPC cluster (University of Queensland).
Inference is performed with temperature set to~0.
For the cross-model analysis
(Section~\ref{sec:overall}), Llama~3.1 8B is run
under the same configuration.
All models are quantised to 4-bit precision (Q4\_K\_M).
Sentence embeddings are computed using
\texttt{all-MiniLM-L6-v2} on the same GPU nodes.
For the RAG setting, retrieval is performed with $k{=}3$
nearest windows and a lookback of 500~prior windows,
using cosine similarity over the computed embeddings.
All dataset-specific choices (the feature schema, the
window size, and the benign-context span) are set a priori
from the documented characteristics of each threat type rather
than tuned on the evaluation labels; the framework therefore
remains zero-shot, using no labelled training data at any stage.
 

\subsection{Baselines}
\label{sec:baselines}
We compare the proposed framework against the Generative
Agent-Based Modelling (GABM) method of
\citet{ferraro2025generative}, a recent LLM-based multi-agent
framework for insider threat detection evaluated on the same
two benchmark datasets used in this work.
GABM employs specialised LLM agents for each log type, whose
analyses are synthesised by a supervisor agent for final
classification using LLaMA-3.1-8B. It classifies per activity identifier rather than per fixed-length
window, and does not specify the procedure mapping ground truth to
labelled instances or how its benign class is sampled (constrained to
5{,}000 of 537{,}840 entries on PicoDomain).
GABM achieves perfect recall on both datasets but suffers from
low precision, indicating a high false positive rate.
Reported results from the original publication are used for
comparison.
GABM uses LLaMA-3.1-8B as its base model, whereas our primary
configuration uses the larger DeepSeek-R1-Distill-Qwen 32B. To
isolate the contribution of the architecture from that of model
scale, we hold the base model fixed at Llama~3.1 8B, the same
model GABM uses. Applied directly to logs without the dual-LLM
architecture, Llama~3.1 8B performs well below the framework
(Section~\ref{sec:direct_llm}); embedded in the proposed framework, the same model performs
substantially better (Section~\ref{sec:overall}), indicating that the
dual-LLM architecture, not model capacity, drives the difference.

\citet{song2024audit} is a related multi-agent LLM approach for insider threat detection but is not used as a direct numerical baseline; we benchmark against GABM as the most directly comparable LLM-based method evaluated on these two datasets, and discuss Audit-LLM as related work.

\subsection{Evaluation Metrics}
\label{sec:metrics}
 
Performance is evaluated using precision, recall, and F1-score. Metrics are computed by aggregating predictions across all users within each dataset (5~users for PicoDomain and 99~users for CERT~r5.2), producing a single overall evaluation per dataset. The F1-score provides a balanced measure of precision and recall, which is the primary basis for comparison in this work.

\subsection{Preliminary Experiment: Direct LLM Classification}
\label{sec:direct_llm}

\begin{figure*}[t]
  \centering
\includegraphics[width=0.75\textwidth]{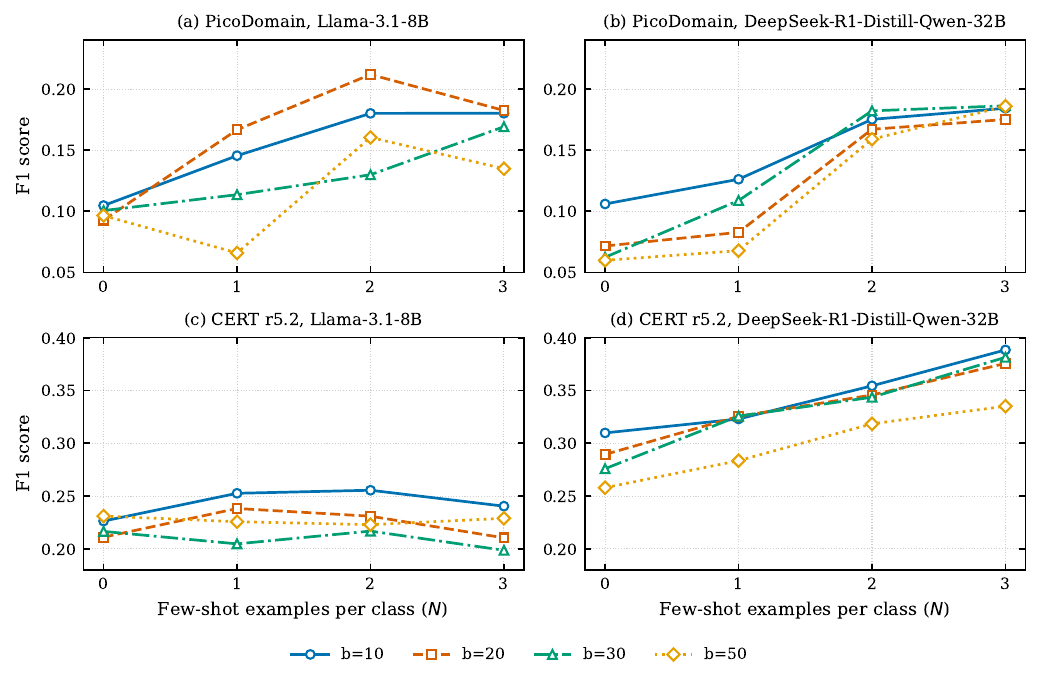}  \caption{Direct LLM classification (reasoning and
classification in a single pass, without windowing, retrieval,
or a dedicated feature-generation stage): F1 against the number
of few-shot examples per class $N$, by batch size $b$, for
(a,~b)~PicoDomain and (c,~d)~CERT~r5.2, using Llama-3.1-8B and
DeepSeek-R1-Distill-Qwen-32B. All configurations remain well
below the proposed framework.}
\label{fig:direct_llm}
\end{figure*}

The state-of-the-art baseline (GABM) decomposes detection across
multiple LLM agents, separating per-log-type reasoning from a
supervisor classification stage. Before adopting a decomposition of
our own, we test whether it can be avoided: whether a single LLM can
carry reasoning and classification in one pass. The model receives
individual raw log records, with all label and ground-truth
annotation fields removed, and must reason over each event and emit a
\textsc{benign}/\textsc{malicious} decision within the same prompt,
without windowing, retrieval, or a dedicated feature-generation stage
(LLM\textsubscript{1}). We sweep this design on both datasets over the
number of few-shot examples per class $N\in\{0,1,2,3\}$ and the batch
size $b\in\{10,20,30,50\}$, using both DeepSeek-R1-Distill-Qwen 32B
and Llama 3.1 8B. Results are shown in Figure~\ref{fig:direct_llm}.

Additional supervision yields only small and inconsistent gains.
Llama 3.1 8B improves up to two-shot and then declines, while
DeepSeek improves slowly with shot count on CERT~r5.2. No
combination of shot count and batch size lifts direct
classification past a low ceiling. At matched zero-shot
supervision, the setting in which the framework itself
operates, the best direct configuration reaches an F1 of only
30.98\% (DeepSeek-R1-Distill-Qwen 32B) and 23.10\% (Llama 3.1
8B) on CERT~r5.2, and 10.60\% and 10.47\% on PicoDomain, all far
below the framework's best of 64.14\% and 50.87\%
(Table~\ref{tab:cross_model}). Because the same models are used
in both settings, this gap reflects the architectural design
rather than model capacity or supervision level.

Across both models and datasets, a single LLM cannot perform the
task end-to-end, and additional prompting does not close the
gap. We attribute this to the demands of a single prompt, which
must parse heterogeneous log fields, reason about threat
behaviour, and commit to a decision at once. This motivates the
structured, multi-stage design of the proposed framework, in
which a feature-generation stage (LLM\textsubscript{1}) produces
an intermediate representation that a separate sequence
classifier (LLM\textsubscript{2}) consumes; the contribution of
retrieval and of the model assigned to each stage is analysed in
Section~\ref{sec:overall}, and the remaining design choices in
the ablation study (Section~\ref{sec:ablation}).


\section{Results and Discussion}\label{sec:results}

\subsection{Classification Performance}
\label{sec:overall}
To examine how model
capability and retrieval interact across the two stages, we
evaluate four model combinations:
(i)~DeepSeek as both feature generator and classifier;
(ii)~Llama~3.1~8B as both;
(iii)~Llama~3.1~8B as feature generator with DeepSeek as
classifier; and
(iv)~DeepSeek as feature generator with Llama~3.1~8B as
classifier. Each configuration is evaluated under both the
RAG and No-RAG settings defined above, yielding eight conditions
per dataset (16 across the two datasets). This cross-model analysis
forms a central investigation of the framework
(Section~\ref{sec:overall}).

The overall performance of the framework across the four model combinations is presented in Table~\ref{tab:cross_model}. Two patterns are consistent across datasets. First, the choice of feature-generating
model (LLM\textsubscript{1}) has a larger effect on
performance than the choice of classifier (LLM\textsubscript{2}):
swapping LLM\textsubscript{2} while holding LLM\textsubscript{1}
fixed moves F1 by less than swapping LLM\textsubscript{1}. Second, the value of retrieval is conditional on feature-generator capacity. When LLM\textsubscript{1} is the weaker model (Llama~3.1 8B), retrieval supplies the comparative context it cannot generate internally, lifting F1 by up to 13.4 percentage points on PicoDomain; when LLM\textsubscript{1} is the stronger model (DeepSeek-R1 32B), retrieval is neutral or slightly negative.

Beyond F1, retrieval shifts the operating point. On CERT~r5.2
under the DeepSeek/DeepSeek combination, the No-RAG configuration
attains its higher F1 through a recall-biased regime: it
produces 4{,}490 false positives versus 3{,}447 with retrieval,
and correctly identifies only 671 of 5{,}161 benign windows
(13.0\% specificity) versus 1{,}714 (33.2\%) with retrieval.
Retrieved per-user behavioural context thus calibrates
LLM\textsubscript{1} risk scores against historical baselines,
yielding a more selective detector. The false-positive
discipline is what distinguishes the framework from the
perfect-recall, low-precision GABM baseline.

Table~\ref{tab:overall} compares the highest-F1 configuration per
dataset against the GABM baseline. On PicoDomain, the framework reaches an F1-score of
50.87\% (DeepSeek as LLM\textsubscript{1}, Llama~3.1 8B as
LLM\textsubscript{2}, No-RAG), an improvement of 31.50
percentage points over the GABM baseline (F1\,=\,19.37\%).
While the baseline attains perfect recall, it does so at the
cost of extremely low precision (10.72\%). The framework
provides a substantially more balanced trade-off, improving
precision by 25.85 percentage points while maintaining a recall
of 83.50\%.

On CERT~r5.2, the framework attains an F1-score of 64.14\%
(Llama~3.1 8B as LLM\textsubscript{1}, DeepSeek as
LLM\textsubscript{2}, RAG), with precision 48.55\% and recall
94.47\%. To verify that this reflects genuine detection rather than
the positive-heavy bias that F1 can reward on enriched evaluation
sets, we report the Matthews correlation coefficient (MCC), which is
zero for any trivial constant classifier. The proposed configuration
attains MCC\,=\,0.146 ($\chi^2\approx204$, $p<10^{-10}$), confirming a statistically significant association between predictions and labels. This pattern is consistent across the configurations in
Table~\ref{tab:cross_model}, indicating that it stems from the
framework's architecture rather than from any single model choice.

Across both datasets, the proposed framework consistently outperforms the GABM baseline on F1 and precision, offering a more balanced precision-recall operating point. The dominant driver of these gains is the structured feature
representation produced by LLM\textsubscript{1}, supported by per-user behavioural grounding; LLM\textsubscript{2} aggregates these per-window indicators into window-level decisions. The remaining design decisions are isolated in the ablation study
(Section~\ref{sec:ablation}).
 
While high recall is desirable in security operations to minimise
missed threats, the framework's higher precision than GABM's reported
results indicates better discrimination between benign and malicious
windows. We do not claim a specific operational false-positive rate:
because this distribution is enriched for malicious activity, precision
on a realistic, predominantly benign stream would be lower
(Section~\ref{sec:limitations}).

\begin{table}[!t]
\centering
\caption{Detection performance (F1-score, \%) across the four
LLM\textsubscript{1}/LLM\textsubscript{2} model combinations, under
retrieval (RAG) and no retrieval (No-RAG), on both datasets.
Rows are grouped by the feature-generating model
(LLM\textsubscript{1}); best F1 per dataset in \textbf{bold}.}
\label{tab:cross_model}
\resizebox{\columnwidth}{!}{%
\begin{tabular}{lcccc}
\toprule
& \multicolumn{2}{c}{CERT~r5.2} & \multicolumn{2}{c}{PicoDomain} \\
\cmidrule(lr){2-3}\cmidrule(lr){4-5}
LLM\textsubscript{1} / LLM\textsubscript{2}
   & RAG & No-RAG & RAG & No-RAG \\
\midrule
DeepSeek / DeepSeek & 58.36 & 61.72 & 45.74 & 46.08 \\
DeepSeek / Llama~8B & 60.53 & 62.31 & 50.17 & \textbf{50.87} \\
\midrule
Llama~8B / Llama~8B & 63.88 & 60.88 & 48.18 & 34.73 \\
Llama~8B / DeepSeek & \textbf{64.14} & 60.08 & 47.24 & 34.85 \\
\bottomrule
\end{tabular}%
}
\end{table}


\begin{table}[!t]
\centering
\caption{Comparison of the proposed framework against the prior
LLM-based baseline (\%). The proposed row reports the best
configuration per dataset (PicoDomain: DeepSeek/Llama~8B, No-RAG;
CERT~r5.2: Llama~8B/DeepSeek, RAG). Best F1 per dataset in
\textbf{bold}.}
\label{tab:overall}
\resizebox{\columnwidth}{!}{%
\begin{tabular}{llccc}
\toprule
Dataset & Method & Precision & Recall & F1 \\
\midrule
\multirow{2}{*}{PicoDomain}
 & GABM~\citep{ferraro2025generative}
   & 10.72 & 100.00 & 19.37 \\
 & Proposed framework
   & 36.57 & 83.50 & \textbf{50.87} \\
\midrule
\multirow{2}{*}{CERT~r5.2}
 & GABM~\citep{ferraro2025generative}
   & 35.82 & 100.00 & 52.74 \\
 & Proposed framework
   & 48.55 & 94.47 & \textbf{64.14} \\
\bottomrule
\end{tabular}%
}
\end{table}

\subsection{ Feature Analysis}
\label{sec:feature_quality}

The cross-model results in Table~\ref{tab:cross_model} show
that the stage at which model capability matters most differs
by dataset: on CERT~r5.2 the strongest configuration places the
larger model at classification (Llama~8B\,/\,DeepSeek, RAG,
F1\, =\,64.14\%), whereas on the stealthier PicoDomain logs it is
required at feature generation (DeepSeek\,/\,Llama~8B, No-RAG,
F1\,=\,50.87\%). To explain this, we examine the discriminative
signal carried by the LLM\textsubscript{1} feature vectors,
using SHAP feature importance~\cite{lundberg2017unified} and
distribution analysis, for DeepSeek-R1-Distill-Qwen 32B and
Llama~3.1 8B on each dataset. The Random Forest underlying the SHAP analysis is used solely as a probe of the intrinsic discriminative content of the LLM\textsubscript{1} feature vectors; it is not part of the proposed pipeline, and the resulting importances characterise the feature set rather than the behaviour of LLM\textsubscript{2}.

\begin{figure*}[!t]
    \centering
    \includegraphics[width=\textwidth]{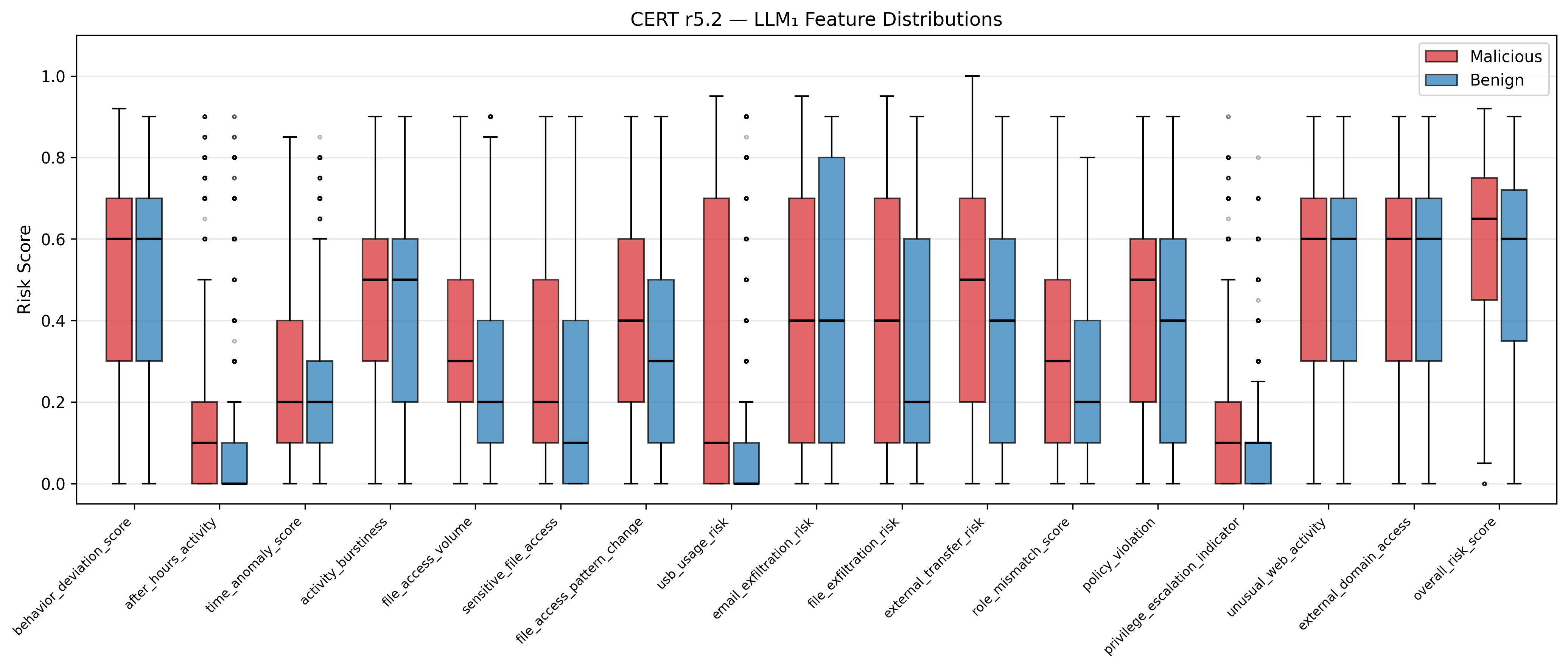}\\
    \includegraphics[width=\textwidth]{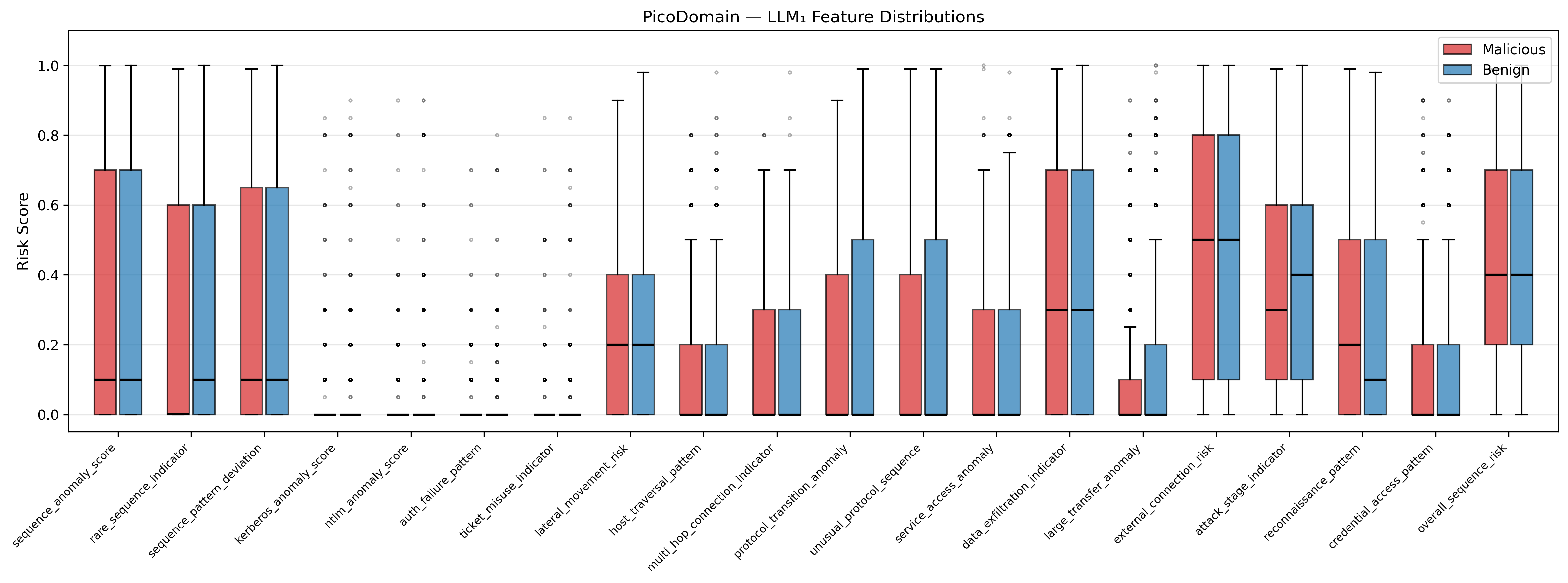}
    \caption{LLM\textsubscript{1} feature score distributions (malicious vs
    benign) for CERT~r5.2 (top) and PicoDomain (bottom). On CERT~r5.2, several
    features show clear separation (notably \textit{usb\_usage\_risk},
    \textit{file\_exfiltration\_risk}, and
    \textit{file\_access\_pattern\_change}), consistent with known insider
    threat behaviours. On PicoDomain, the malicious and benign distributions of
    the LLM\textsubscript{1} scores are near-identical across all 20 features.}
\label{fig:feature_dist}
\end{figure*}

\begin{figure*}[!t]
    \centering
    \includegraphics[width=\textwidth]{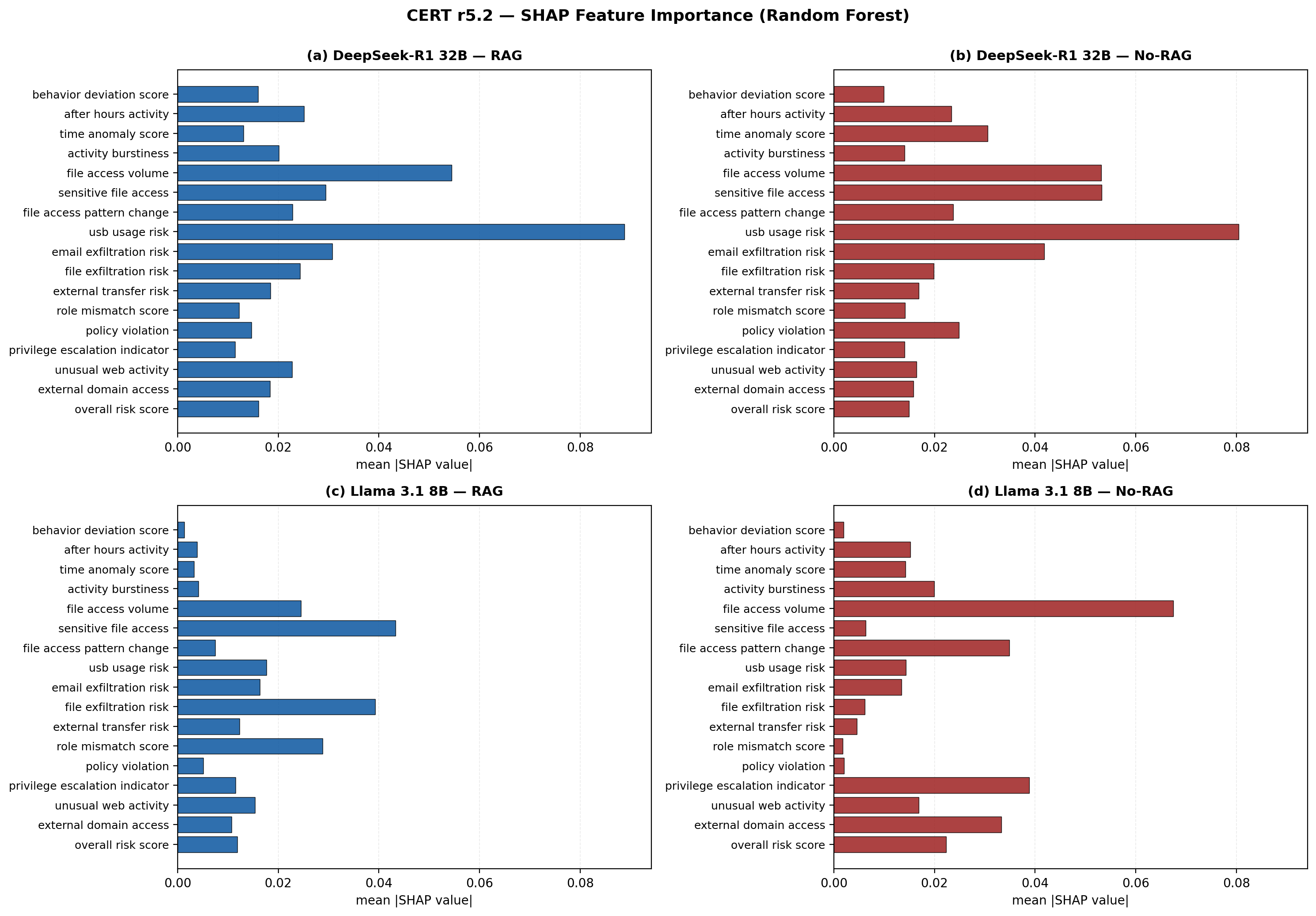}
    \caption{LLM\textsubscript{1} feature importance on CERT~r5.2:
    mean $|$SHAP$|$ from TreeSHAP~\citep{lundberg2017unified} over a
    Random Forest trained on the LLM\textsubscript{1} feature
    vectors, for both feature-generation models under RAG and
    No-RAG. Features appear in extraction order with a shared
    x-axis across all four panels.}
    \label{fig:shap_cert}
\end{figure*}

\paragraph{CERT~r5.2.}
To validate LLM\textsubscript{1} feature quality on CERT ~r5.2, we analyse
feature score distributions (Figure~\ref{fig:feature_dist}) and SHAP feature
importance (Figure~\ref{fig:shap_cert}). DeepSeek-R1 32B ranks
\textit{usb\_usage\_risk} as the dominant contributor (20.2\% of total
importance), consistent with USB-based exfiltration scenarios in CERT~r5.2.
Llama~3.1 8B produces a different ranking, prioritising
\textit{sensitive\_file\_access} (16.9\%) and \textit{file\_exfiltration\_risk}
(15.3\%). Although the single top-ranked feature differs, both models
concentrate importance on the same family of file-access and exfiltration
indicators, the features that also show the clearest malicious/benign
separation in Figure~\ref{fig:feature_dist}. The separation is substantial:
the mean absolute difference between class averages across the 17~features is
0.057, with \textit{usb\_usage\_risk} reaching 0.210 (malicious mean 0.295
versus benign 0.085). The agreement between the two analyses indicates the
discriminative signal is robust to model choice even where the exact ranking
is not. Under No-RAG, DeepSeek's anomaly features sharpen relative to RAG
(e.g.\ \textit{sensitive\_file\_access}, \textit{time\_anomaly\_score}),
indicating that retrieval redistributes feature importance rather than adding
discriminative power, consistent with the substitution pattern in
Table~\ref{tab:cross_model}.

\begin{figure*}[!t]
    \centering
    \includegraphics[width=\textwidth]{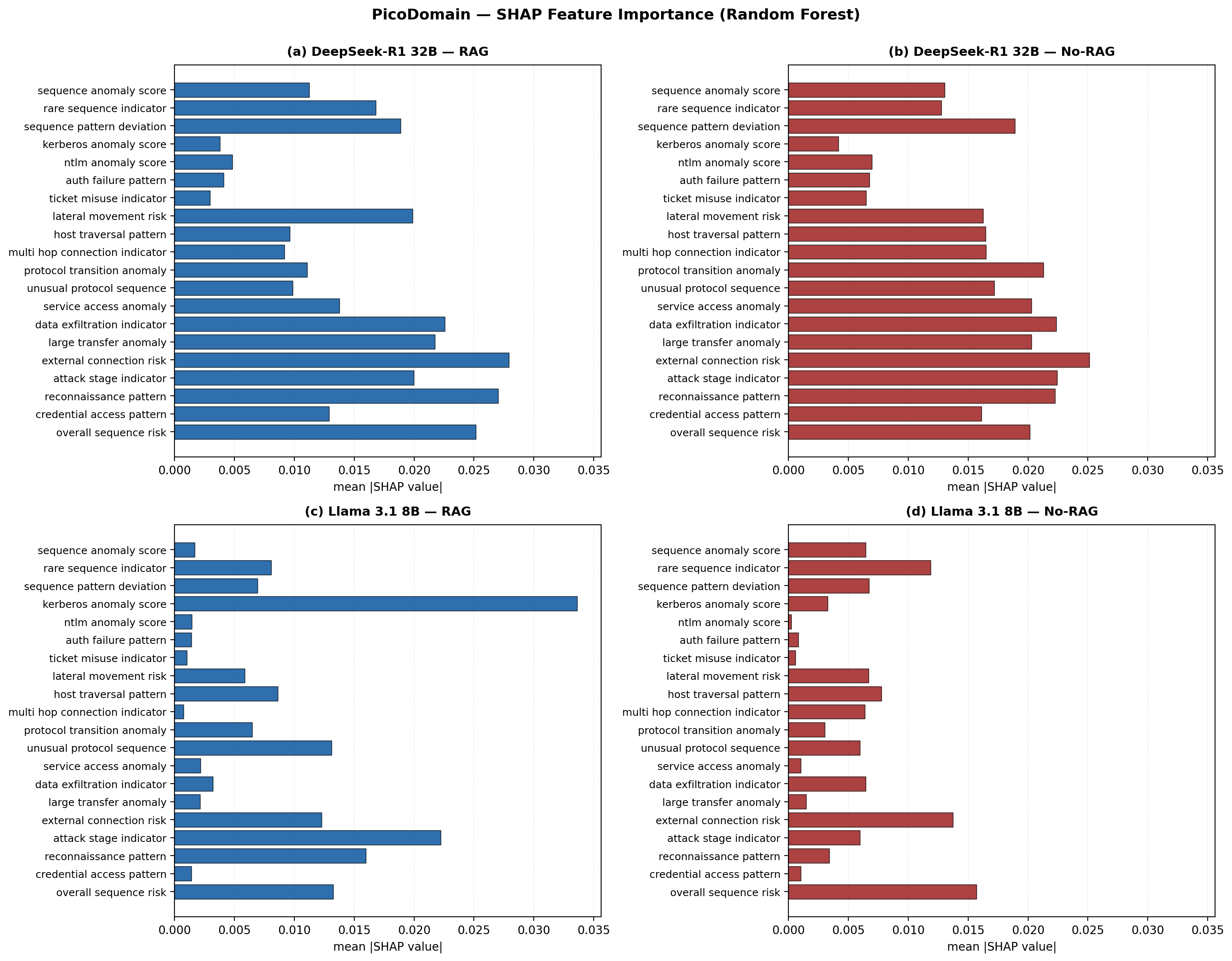}
    \caption{LLM\textsubscript{1} feature importance on PicoDomain;
    same method and layout as Figure~\ref{fig:shap_cert}.}
    \label{fig:shap_pico}
\end{figure*}

\paragraph{PicoDomain.}
On PicoDomain, the LLM\textsubscript{1} risk scores show very weak
malicious/benign separation: the mean absolute difference between class
averages across the 20~indicators is 0.008 (maximum 0.030), roughly seven-fold
smaller than on CERT~r5.2; per-window feature variance is essentially identical
between classes (0.043 vs 0.044 for DeepSeek-R1 32B); and both classes activate
a near-identical number of features per window (9.4 vs 9.5 of 20). The largest
differences are moreover mildly inverted: benign windows score \emph{higher}
than malicious ones (e.g.\ \textit{large\_transfer\_anomaly}, 0.130 vs 0.100), reflecting that stealthy beaconing produces per-window records less
conspicuous than ordinary traffic, a characteristic of the threat rather than a
failure of feature design. SHAP analysis (Figure~\ref{fig:shap_pico}) shows the
same pattern, with maximum mean $|$SHAP$|$ below 0.035, roughly threefold
smaller than the dominant CERT~r5.2 features ($\approx$0.089). The two models
also differ in where they place importance: DeepSeek-R1 32B spreads it across
network-level indicators (\textit{external\_connection\_risk},
\textit{reconnaissance\_pattern}, \textit{overall\_sequence\_risk}) with no
single dominant feature, whereas Llama~3.1 8B under RAG concentrates on
\textit{kerberos\_anomaly\_score} (a feature that shows no malicious/benign
separation in Figure~\ref{fig:feature_dist}), indicating reliance on a
non-discriminative signal. This weak per-window signal (Figure~\ref{fig:shap_pico}) is reflected in the
prevalence-robust metric: on PicoDomain the best configuration attains
MCC\,=\,0.004, statistically indistinguishable from chance, in contrast to the
significant association on CERT~r5.2 (MCC\,=\,0.146). We therefore treat
per-window detection on PicoDomain as a boundary case for this class of
method, attributable to the near-absence of extractable signal in stealthy
beaconing traffic rather than to a failure of feature design, and not as
evidence of reliable detection.

\subsection{Per-User Classification Performance}
\label{sec:per_user}

Table~\ref{tab:peruser_pico} presents per-user results on
PicoDomain for the best configuration (DeepSeek as
LLM\textsubscript{1}, Llama~3.1 8B as LLM\textsubscript{2},
No-RAG). Performance varies considerably across users, with
F1-scores ranging from 40.00\% to 61.37\%. Per-user results for
CERT~r5.2 are omitted due to the large number of users (99);
combined results for all configurations are reported in
Table~\ref{tab:cross_model}.

Recall is uniformly high across the four larger users
(79.65\%--86.82\%), indicating the framework detects most
malicious windows regardless of user; precision is the binding
constraint throughout (29.30\%--47.46\%). JSNAKE achieves the
strongest result (F1\,=\,61.37\%), combining the highest
precision (47.46\%) with high recall, followed by BDUCK
(52.99\%) and RMOLE (52.08\%). JDOE records the lowest precision
(29.30\%): its traffic is predominantly benign (520 of 751
windows), so false positives accumulate disproportionately.
ADMINISTRATOR scores lowest overall (40.00\%) on a very small
sample (11 malicious and 20 benign windows), where individual
misclassifications shift the score considerably.

\begin{table}[!t]
\centering
\caption{Per-user results on PicoDomain for the best
         configuration (DeepSeek/Llama~8B, No-RAG).}
\label{tab:peruser_pico}
\resizebox{\columnwidth}{!}{%
\begin{tabular}{lcccrr}
\toprule
User & Prec. & Recall & F1 & Mal. & Ben. \\
\midrule
ADMINISTRATOR & 31.58 & 54.55 & 40.00 &  11 &  20 \\
BDUCK         & 38.96 & 82.82 & 52.99 & 326 & 513 \\
JDOE          & 29.30 & 79.65 & 42.84 & 231 & 520 \\
JSNAKE        & 47.46 & 86.82 & 61.37 & 129 & 158 \\
RMOLE         & 37.20 & 86.80 & 52.08 & 303 & 530 \\
\midrule
\textbf{Combined}
  & \textbf{36.57} & \textbf{83.50}
  & \textbf{50.87}
  & \textbf{1{,}000} & \textbf{1{,}741} \\
\bottomrule
\end{tabular}%
}
\end{table}


\subsection{Ablation Study}
\label{sec:ablation}

We analyse the contribution of two design decisions: the window
representation fed to LLM\textsubscript{1}, and the log-grouping
strategy. The window-representation ablation is conducted on
CERT~r5.2, whose five log sources carry rich and varied per-event
fields; how each window is encoded materially affects the signal
available to LLM\textsubscript{1}. PicoDomain's discriminative signal is instead concentrated in the protocol and destination of each connection, already captured by the sequence representation, and a comparable representation study is therefore left for future work. The
grouping-strategy ablation is conducted on PicoDomain because
connection-pair grouping is a network-specific alternative with no
counterpart in CERT~r5.2's host activity logs, where per-user
grouping instead serves to merge the five log sources into a single
timeline. The contribution of retrieval and of the model assigned to
each stage is analysed jointly in Section~\ref{sec:overall}
(Table~\ref{tab:cross_model}).

\subsubsection{Effect of window representation}
Table~\ref{tab:ablation_repr} isolates the effect of window
representation on CERT~r5.2, holding the model, retrieval method, and
feature set constant (DeepSeek-R1-32B, embedding RAG, 17 features). We compare
three representations of the same underlying log window:
(i)~\emph{verbose event text}: the raw natural-language log
lines of the window concatenated into a single text block;
(ii)~\emph{sequence + destination tokens}: a compact summary
listing the ordered sequence of activity types and a
deduplicated set of all destinations referenced (recipients,
URLs, file paths, hosts, attachment names); and
(iii)~\emph{structured event objects}: each log entry
rendered as a JSON object retaining the fields relevant to
threat reasoning (timestamp, action, sender, recipients, URL,
filename, file tree, host, size, attachments, and USB flags),
with events preserved in temporal order.

Verbose full-text representation underperforms the compact
sequence-and-destination format, which is in turn outperformed
by structured event objects, the representation adopted in the
proposed system. Structured, temporally rich input provides the
most discriminative signal for LLM\textsubscript{1} feature
extraction and is the dominant controllable factor in detection
performance on CERT~r5.2.

\begin{table}[!t]
\centering
\caption{Effect of window representation on CERT~r5.2. All
         configurations use DeepSeek-R1-32B with embedding RAG and 17
         dataset-specific features.}
\label{tab:ablation_repr}
\resizebox{\columnwidth}{!}{%
\begin{tabular}{lccc}
\toprule
Window Representation & Prec.~(\%) & Recall~(\%) & F1~(\%) \\
\midrule
Verbose event text & 45.84 & 62.52 & 52.89 \\
Sequence + destination tokens & 47.41 & 66.83 & 55.47 \\
Structured event objects (proposed) & 48.43 & 73.40 & \textbf{58.36} \\
\bottomrule
\end{tabular}%
}
\end{table}

\subsubsection{Effect of grouping strategy}
A core component of the framework is per-user log grouping
(Section~\ref{sec:methodology}), which merges all log entries
associated with a single user into a unified chronological
timeline. To validate this design, we compare it against
connection-pair grouping on PicoDomain, in which windows are
constructed from logs grouped by source--destination host pair
rather than by user. Here each unique pair forms its own log
sequence, segmented into windows by the same sliding-window
procedure used for per-user grouping. Both configurations use
identical features (20 network indicators), DeepSeek-R1-32B with
embedding RAG, and inference settings; only the grouping unit
differs.
As shown in Table~\ref{tab:ablation_grouping}, per-user
grouping substantially outperforms connection-pair grouping
(F1\,=\,45.74\% vs 29.87\%), with the largest gain in precision
(36.16\% vs 20.45\%). Connection-pair grouping produces three
times more windows (8{,}320 vs 2{,}741) by fragmenting each
user's activity across 83 connection pairs, diluting the
behavioural context available to LLM\textsubscript{1}.
User-level timelines instead preserve the full scope of each
user's activity across all connection types, providing more
coherent context for threat-aware feature extraction.

\begin{table}[!t]
\centering
\caption{Effect of log-grouping strategy on PicoDomain. Both
configurations use identical features (20 network indicators),
DeepSeek-R1-32B with embedding RAG, and inference settings; only the
grouping unit differs.}
\label{tab:ablation_grouping}
\resizebox{\columnwidth}{!}{%
\begin{tabular}{lccc}
\toprule
Grouping & Prec.~(\%) & Recall~(\%) & F1~(\%) \\
\midrule
Connection-pair (source--destination) & 20.45 & 55.41 & 29.87 \\
Per-user (proposed) & 36.16 & 62.20 & \textbf{45.74} \\
\bottomrule
\end{tabular}%
}
\end{table}


\subsection{Limitations}
\label{sec:limitations}

First, the temporal-context sampling evaluates attack-proximal windows, not
a continuous stream. The subset is enriched for malicious activity, so the
reported precision reflects this distribution rather than an operational
false-positive rate, which would be lower on a predominantly benign stream.
As malicious windows appear in bursts, part of LLM\textsubscript{2}'s
performance may derive from this structure rather than per-window content.

Second, LLM\textsubscript{1} features carry almost no per-window signal on
PicoDomain, and detection there is at chance (MCC\,=\,0.004) against a
significant association on CERT~r5.2 (MCC\,=\,0.146). We treat PicoDomain as
a boundary case for this class of method, reflecting stealthy beaconing
rather than a feature-design flaw (Section~\ref{sec:feature_quality}).

Third, we do not compare absolute metrics against GABM: it classifies per
activity identifier rather than per window, and specifies neither its
labelling nor its benign-sampling procedure (5{,}000 of 537{,}840 entries,
a prevalence near 1.6\% against our 36.5\%), so a like-for-like comparison
is not definable. We rely instead on prevalence-robust metrics and internal
comparisons on an identical subset. Generalisation beyond these two datasets,
and to other prompts and embedding models, remains to be established.
 
\section{Conclusion}\label{sec:conclusion}
We presented a retrieval-augmented dual-LLM framework for
zero-shot intrusion detection across heterogeneous security
logs. Logs are grouped into per-user chronological timelines
against which each window is assessed. LLM\textsubscript{1}
transforms raw log windows into structured vectors of
interpretable risk indicators tailored to the threat type, and
LLM\textsubscript{2} classifies ordered sequences of these
vectors to detect multi-window attack patterns.

Evaluated on CERT~r5.2 and PicoDomain, the framework achieves
best F1-scores of 64.14\% and 50.87\% respectively, improvements of 11.40 and 31.50 percentage points over the GABM
baseline, and every one of the eight evaluated
configurations exceeds the baseline on both datasets,
indicating the gains stem from the architecture rather than any
single model choice. The cross-model analysis suggests two patterns. First, retrieval and model capability appear to act as
substitutes: retrieval improves the features produced by the
less capable model, especially on weak-signal data, whereas the
more capable model attains comparable feature quality without
retrieved context. Second, the best configurations assign
different models to the two stages, placing the more capable
model at feature generation on weak-signal data and at
classification on strong-signal data.

Feature analysis shows that LLM\textsubscript{1} feature
quality reflects the nature of each threat type: insider threat
activity on CERT~r5.2 produces clear behavioural traces (mean
absolute class difference 0.057 across the 17~features), while
APT beaconing on PicoDomain is inherently stealthy (0.008,
with the largest differences mildly inverted). The structured representation nonetheless provides a consistent basis for LLM\textsubscript{2}'s window-level classification, despite the limited discriminative signal at the individual window level.

Future work includes evaluating on additional datasets,
investigating cross-user retrieval to surface shared C2
beaconing patterns across hosts, and exploring fine-tuning LLMs
for the feature generation stage.



\bibliographystyle{cas-model2-names}
\bibliography{cas-refs}
\end{document}